\documentclass[a4paper,fleqn,usenatbib]{mnras}

\usepackage[T1]{fontenc}
\usepackage{ae,aecompl}
\usepackage{graphicx}	
\usepackage{natbib}
\usepackage{amsmath}	
\usepackage{amssymb}	
\usepackage{txfonts}

\title[New bright magnetic B stars]{Discovery of two new bright magnetic B
stars: i\,Car and Atlas\thanks{Based on observations obtained at the T\'elescope
Bernard Lyot (USR5026) operated by the Observatoire Midi-Pyr\'en\'ees,
Universit\'e de Toulouse (Paul Sabatier), Centre National de la Recherche
Scientifique (CNRS) of France, and at the European Southern Observatory (ESO),
Chile (program ID 094.D-0274B).}}

\author[C. Neiner et al.]{
Coralie Neiner$^{1}$\thanks{E-mail: coralie.neiner@obspm.fr},
Bram Buysschaert$^{1,2}$,
Mary E. Oksala$^{1}$
and Aurore Blazere$^{1,3}$
\\
$^{1}$LESIA, Observatoire de Paris, PSL Research University, CNRS,
Sorbonne Universit\'es, UPMC Univ. Paris 06, Univ. Paris Diderot,\\
Sorbonne Paris Cit\'e, 5 place Jules Janssen, 92195 Meudon, France\\
$^{2}$Instituut voor Sterrenkunde, KU Leuven, Celestijnenlaan 200D, 3001 Leuven, Belgium\\
$^{3}$Universit\'e de Toulouse, UPS-OMP, IRAP, CNRS, 14 avenue Edouard Belin, 31400 Toulouse, France
}

\date{Accepted XXX. Received YYY; in original form ZZZ}
\pubyear{2015}

\begin{document}
\label{firstpage}
\pagerange{\pageref{firstpage}--\pageref{lastpage}}
\maketitle

\begin{abstract}
The BRITE (BRIght Target Explorer) constellation of nano-satellites performs
seismology of bright stars via high precision photometry. In this context, we
initiated a high resolution, high signal-to-noise, high sensitivity,
spectropolarimetric survey of all stars brighter than V=4. The goal of this
survey is to detect new bright magnetic stars and provide prime targets for both
detailed magnetic studies and asteroseismology with BRITE. Circularly polarised
spectra were acquired with Narval at TBL (France) and HarpsPol at ESO in La
Silla (Chile). We discovered two new magnetic B stars: the B3V star i\,Car and
the B8V component of the binary star Atlas. Each star was observed twice to
confirm the magnetic detections and check for variability. These bright magnetic
B stars are prime targets for asteroseismology and for flux-demanding
techniques, such as interferometry.
\end{abstract}



\section{Introduction}

The BRITE (BRIght Target Explorer) constellation of nano-satellites
\citep{weiss2014} photometrically monitors the variations of stars with V$\le$4,
with high precision and cadence, in order to perform asteroseismology. The
mission consists of 3 pairs of nano-satellites, built by Austria, Canada, and
Poland, carrying 3-cm aperture telescopes. One instrument per pair is equipped
with a blue filter, the other with a red filter. Presently, six nano-satellites
are flying, and five are observing. Each BRITE nano-satellite can observe up to
$\sim$25 bright stars, as well as additional fainter targets with reduced
precision. 

Since the BRITE sample consists of the brightest stars, it is dominated by the
most intrinsically luminous stars: hot stars at all evolutionary stages, and
evolved cooler stars (cool giants and AGB stars). In particular, analysis of OB
star variability will help to investigate outstanding issues in stellar physics,
such as the sizes of their convective cores, their internal rotation profiles,
and the influence of rapid rotation on their structure and evolution. Several
types of pulsating hot stars are known. Pulsations in O stars are excited by the
$\kappa$ mechanism, and red noise observed in their power spectra
\citep{blomme2011} has been interpreted as stochastically excited internal waves
\citep{aerts2015}. B stars also undergo pulsations excited by the $\kappa$
mechanism: the early types (B0-B2) are $\beta$\,Cep pulsators exhibiting mostly
pressure modes \citep{dziembowski1993_bc}, while the later types (B2-B9) are
Slowly Pulsating B stars (SPB) with mostly gravity modes
\citep{dziembowski1993_SPB}. Classical Be stars show $\beta$\,Cep and SPB-type
$\kappa$-driven pulsations, but their rapid rotation also leads to the
stochastic excitation of gravito-inertial waves
\citep{neiner2012_stocha,mathis2014,lee2014}. Finally, the pulsations of OB
supergiants are less well understood, but they seem to be driven by the
$\epsilon$ mechanism \citep{moravveji2012,saio2015}, and possibly stochastic
excitation.

In addition to pulsations, about 10\% of all hot stars are found to be magnetic
\citep{grunhutneiner2015}, and the origin of their magnetic field is fossil,
i.e. a descendant of the field present in the molecular cloud from which the
star formed \citep{neiner2015}. Over the last decade, thanks to large
spectropolarimetric surveys of hot stars, such as MiMeS
\citep{wade2014,wade2015}, the number of known magnetic hot stars has
significantly increased. Nevertheless, their number remains low, with specific
types of magnetic hot targets, such as O stars, pulsating B stars, or
supergiants, still relatively poorly studied. Moreover, only about ten known
magnetic massive stars are brighter than V=4. 

The study of the magnetic properties of pulsating hot stars is very interesting
since, when combined with the study of their pulsational properties, it provides
unique information about the interior of hot stars. The combination of an
asteroseismic study with a spectropolarimetric study has been accomplished for
only a few hot stars so far, e.g., for the $\beta$\,Cep star V2052 Oph. This
star presents a magnetic field with a polar strength  of about 400 G
\citep{neiner2012}, which is above the critical field limit needed to inhibit
mixing inside the star as determined from theory \cite[e.g.,][]{zahn2011}.
Asteroseismic models explaining the observed pulsational behaviour of V2052\,Oph
indeed required no convective core overshooting \citep{briquet2012}, contrary
to  non-magnetic $\beta$\,Cep stars \citep[e.g.,][]{briquet2007}. Moreover, in a
pulsating magnetic star, the magnetic field splits the pulsation modes, modifies
their amplitude, or can even inhibit certain modes by redistributing the energy
into other modes. As a consequence, knowing that a field is present, its
strength, and its configuration is essential to properly identify the pulsation
modes and put strong constraints on seismic models. In addition, combining
asteroseismology and magnetism allows us to probe the magnetic field strength
and configuration inside the star, while spectropolarimetric measurements alone
only probe the surface field. 

Therefore, it is very useful to identify bright pulsating magnetic hot stars. In
this frame, we are performing a spectropolarimetric survey of all BRITE targets,
i.e. $\sim$600 stars with V$\le$4, with the goal of discovering new bright
magnetic stars, and thus providing prime targets for BRITE asteroseismic
studies. Spectropolarimetric observations of each of the $\sim$600 BRITE targets
(V$\le$4) are currently being gathered either from archives ($\sim$100 stars) or
with the 3 high-resolution spectropolarimeters available in the world ($\sim$500
stars): Narval at the Bernard Lyot Telescope (TBL) in France, ESPaDOnS at CFHT
in Hawaii, and HarpsPol on the ESO 3.6-m telescope in La Silla. 

In this paper, we present the spectropolarimetric observations of two bright B
stars with Narval and HarpsPol (Sect.~\ref{obs}), and the discovery of their
magnetic field (Sect.~\ref{results}). We then conclude on the great interest of
these bright magnetic stars for further studies (Sect.~\ref{discus}). 

\section{Observations}\label{obs}

The Narval spectropolarimeter covers a wavelength range from about 375 to 1050
nm, with a resolving power of $\sim$68000, spread on 40 echelle orders. The
HarpsPol spectropolarimeter covers a shorter wavelength range from about 380 to
690 nm on two detectors and 71 echelle orders, but with a higher resolving power
of $\sim$110000.

We observed our targets in circular polarisation mode. Each observation consists
of 4 sub-exposures taken in a specific configuration of the polarimeter. The 4
sub-exposures are constructively combined to obtain the Stokes V spectrum in
addition to the intensity (Stokes I) spectrum. The sub-exposures are also
destructively combined to produce a null polarisation (N) spectrum to check for
pollution by, e.g., instrumental effects, variable observing conditions, or
non-magnetic physical effects such as pulsations. In addition, successive Stokes
V sequences can be acquired to increase the total signal-to-noise ratio (S/N) of
a magnetic measurement.

The usual bias, flat-field and ThAr calibrations were obtained each night and
applied to the data. The Narval data reduction was performed using Libre-Esprit
\citep{donati1997}, a dedicated software available at TBL. The HarpsPol
reduction was performed with a modified version of the REDUCE package 
\citep{piskunov2002,makaganiuk2011}. The Stokes I spectra were then normalized
to the continuum level using IRAF\footnote{IRAF is distributed by the National
Optical Astronomy Observatory, which is operated by the Association of
Universities for Research in Astronomy (AURA) under a cooperative agreement with
the National Science Foundation.}, and the same normalization was applied to the
Stokes V and null spectra.

Finally, we applied the Least Squares Deconvolution (LSD) method
\citep{donati1997} to produce a set of LSD Stokes I, Stokes V, and N profiles
for each magnetic measurement. LSD requires a mask listing the lines in the
spectrum, their wavelength, depth, and Land\'e factor. Such a line mask was
produced for each star. We first extracted line lists from the VALD3 atomic
database \citep{piskunov1995, kupka1999} for the appropriate temperature and
gravity of each star. We only used lines (including He lines) with a depth
larger than 0.1. We then removed from the masks all lines that are not visible
in the intensity spectra, hydrogen lines because of their Lorentzian broadening,
those blended with H lines or interstellar lines, as well as lines in regions
affected by absorption of telluric origin. Finally, the depth of each line in
the LSD masks was  adjusted sossibleo as to fit the observed line depth. 

Consecutive sequences were then co-added to produce one magnetic measurement. 

\begin{table}
\caption{Journal of observations indicating the name of the stars, the
instrument used for the spectropolarimetric measurements (H=HarpsPol, N=Narval),
the Heliocentric Julian Date at the middle of the observations (mid-HJD -
2450000), the exposure time, and the mean signal-to-noise ratio of the
(co-added) spectrum at $\sim$500 nm.}
\begin{tabular}{lllllr}
\hline
Star & Instr. & Date & mid-HJD & T$_{\rm exp}$ & S/N \\
\hline
i\,Car	  & H & Mar 3, 2015  & 7084.620 & 5*4*337  & 1272 \\
i\,Car	  & H & Mar 9, 2015  & 7090.712 & 5*4*337  & 1510 \\
Atlas	  & N & Nov 13, 2014 & 6974.600 & ~~~4*245 &  691 \\
Atlas	  & N & Nov 20, 2014 & 6982.515 & 6*4*245  & 3643 \\
\hline
\end{tabular}
\label{tableobs}
\end{table}

\subsection{i\,Car}

i\,Car (HD\,79447, HR\,3663) is a B3V star with magnitude V=3.95. It was 
observed twice with HarpsPol on March 3 and 9, 2015. Each measurement consisted
of 5 consecutive Stokes V sequences of 4 sub-exposures of 337 seconds each, i.e.
a total exposure time of 6740 seconds per magnetic measurement (see
Table~\ref{tableobs}). After applying LSD, the two sets of 5 sequences have been
co-added to produce two magnetic measurements. For the line mask, we started
from a VALD3 line list with $T_{\rm eff}$=18000 K and $\log g$= 3.5, following
parameters available in the literature \citep[e.g.][]{zorec2009,soubiran2010}.
The final mask produced for this star includes 1249 lines. The LSD profiles have
a S/N of 8600 and 8100 in Stokes I, and 24600 and 31100 in Stokes V, for the two
measurements respectively.

\subsection{Atlas}

Atlas (27\,Tau, HD\,23850) is a visual binary system with V=3.63, and a member
of the Pleiades (M45) cluster. \cite{dommanget2000} indicated that the A
component has a magnitude V=3.8, while the B companion has a magnitude V=6.8,
and their separation is 0.4 arcsec. \cite{renson2009} flagged the A component as
a He-weak star, and \cite{wraight2012} provided a tentative variation period of
2.4624 d. The A component was also found to be a close spectroscopic binary
(SB2) system Aa+Ab. Interferometry showed that the Aa and Ab components are
separated by 13 mas \citep{pan2004} in an eccentric (e$\sim$0.24) orbit with a
period of $\sim$291 d \citep{zwahlen2004}. The SB2 system consists of a rapidly
rotating B8III star with a B8V companion \citep{pan2000}, with respective
projected rotational velocities of about $v\sin i$=240 and 60 km~s$^{-1}$.

Atlas has been observed twice with Narval on November 13 and 20, 2014. Since the
diameter of the fiber of Narval is 2.8 arcsec, all 3 components of Atlas were
recorded in the spectra. The first measurement consisted of 2 consecutive Stokes
V sequences of 4*245 seconds. However, the second sequence has a poor
signal-to-noise ratio (S/N) and, as a consequence, only one sequence is used
here. The second measurement consisted of 6 successive sequences of 4*245
seconds in order to improve the S/N. After applying LSD, these 6 sequences have
been co-added to produce one single magnetic measurement. See
Table~\ref{tableobs}.

For the line mask, we started from a VALD3 line list with $T_{\rm eff}$=13000 K
and $\log g$= 3.5, according to the values available in the literature
\citep[e.g.][]{soubiran2010,david2015}. To this template mask, we added missing
\ion{Ne}{i} and \ion{N}{ii} lines extracted from VALD, which are not available
in VALD3. The final mask produced for this star includes 1201 lines. The two LSD
profiles have a S/N of 4673 and 12144 in Stokes I, and 15125 and 74089 in Stokes
V, respectively.

\section{Magnetic analysis and results}\label{results}

\begin{table}
\caption{Longitudinal field ($B_l$) and null (N) measurements in Gauss, with
their respective error bars $\sigma$, significance level $z$, and the magnetic
detection status.}
\begin{tabular}{@{\,\,}llllll@{\,\,}}
\hline
Star & $B_l \pm \sigma B_l$ & $z_B$ & $N \pm \sigma N$ & $z_N$ & Detection \\
\hline
i\,Car	  &   -34.4 $\pm$ 6.8   &  5.1 &  -1.9 $\pm$ 6.8   & 0.3 & Definite \\
i\,Car	  &  -245.8 $\pm$ 5.3   & 46.4 &  -1.3 $\pm$ 5.3   & 0.2 & Definite \\
Atlas	  &   543.6 $\pm$ 127.2 &  4.3 &  -9.8 $\pm$ 127.1 & 0.1 & Definite \\
Atlas	  &   166.4 $\pm$ 19.9  &  8.4 &  41.7 $\pm$ 19.8  & 2.1 & Definite \\
\hline
\end{tabular}
\label{tableresults}
\end{table}

The detection of a magnetic field is evaluated by the False Alarm Probability
(FAP) of a signature in the LSD Stokes V profile inside the LSD line, compared
to the mean noise level in the LSD Stokes V profile outside the line. We adopted
the convention defined by \cite{donati1997}: if FAP $<$ 0.001\%, the magnetic
detection is definite, if 0.001\% $<$ FAP $<$ 0.1\% the detection is marginal,
otherwise there is no magnetic detection. 

\subsection{i\,Car}

\begin{figure}
\resizebox{0.89\hsize}{!}{\includegraphics[clip]{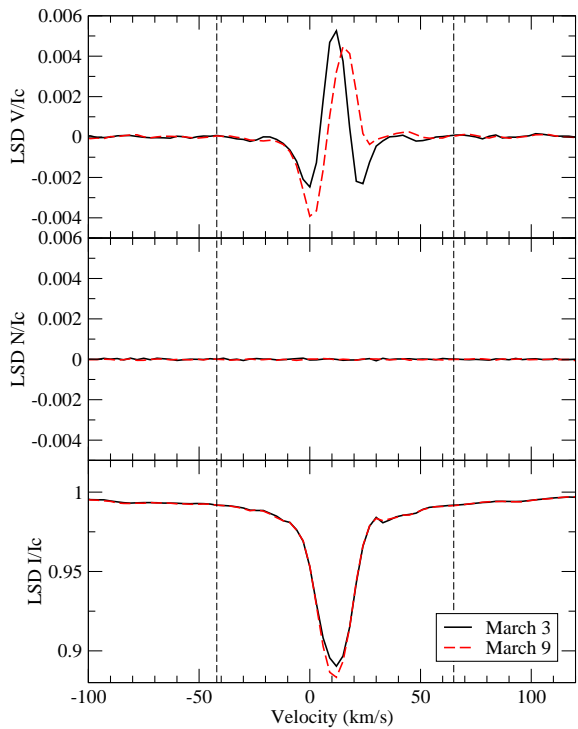}}
\caption[]{LSD Stokes V (top), N (middle) and I (bottom) profiles for the first
(black solid line) and second (red dashed line) measurements of the B3V star
i\,Car. Vertical dashed lines show the integration range for the $B_l$ anf FAP
calculations.}
\label{icar}
\end{figure}

Both LSD Stokes V profiles show definite detections of a magnetic field (with
100\% probability), with a Zeeman signature covering the width of the Stokes I
profile, while the N profiles show no evidence of pollution of the measurements
(see Fig.~\ref{icar}).

Using the center-of-gravity method \citep{rees1979,wade2000} with a mean
wavelength of 500~nm and a mean Land\'e factor of $\sim$1.46, we calculated the
longitudinal field value corresponding to these Zeeman signatures over the
velocity range [-42:65] km~s$^{-1}$. The significance level $z_B = B_l/\sigma
B_l$ of the magnetic measurements is high, while the value for the N
measurements ($z_N$) is very low, confirming that the signature is of stellar
magnetic origin. Results are shown in Table~\ref{tableresults}.

\subsection{Atlas}

\begin{figure}
\resizebox{0.95\hsize}{!}{\includegraphics[clip]{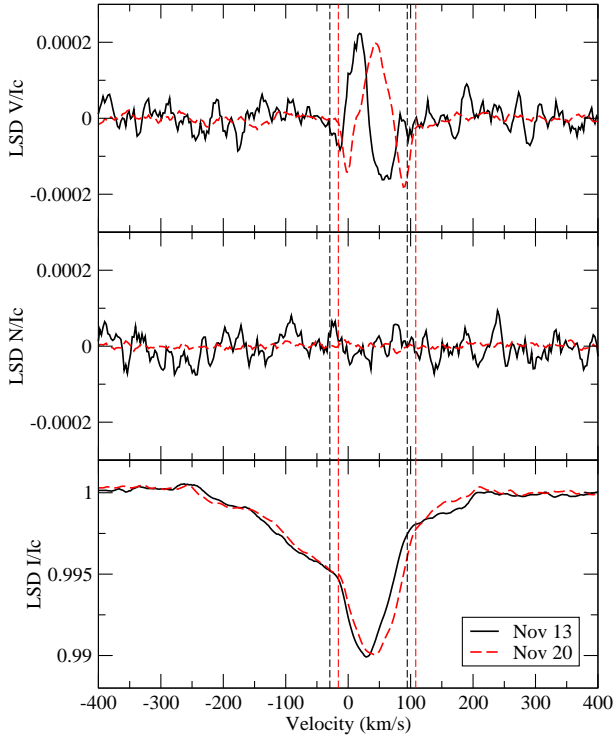}}
\caption[]{LSD Stokes V (top), N (middle) and I (bottom) profiles for the first
(black solid line) and second (red dashed line) B8 multiple star
Atlas. Vertical dashed lines show the integration range for the $B_l$ and FAP
calculations, for the first (black) and second (red) measurements.}
\label{atlas}
\end{figure}

Atlas is a known multiple system, and at least two components are clearly
visible in the LSD Stokes I profiles. Moreover, a slight radial velocity shift
is observed between the two measurements obtained one week apart. The Zeeman
signature covers the width of the Stokes I profile of the narrow-line component,
which is thus the magnetic star. No magnetic signature is observed for the
broad-line component. For the narrow-line component, both LSD Stokes V profiles
show definite detections of a magnetic field (with 100\% probability), while the
N profiles show no evidence of pollution of the measurements (see
Fig.~\ref{atlas}).

\begin{figure}
\resizebox{0.94\hsize}{!}{\includegraphics[clip]{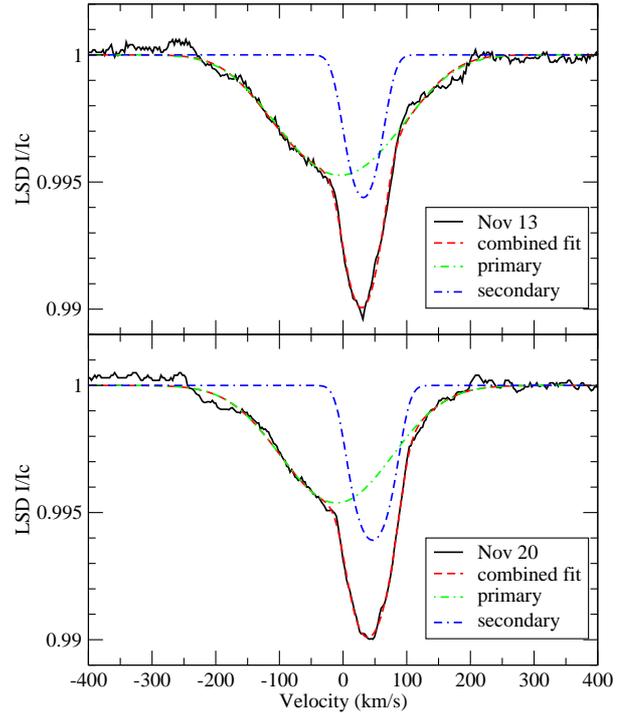}}
\caption[]{LSD Stokes I profiles (black solid line) of the first (top
panel) and second (bottom panel) measurements of the B8III+B8V binary star
Atlas, together with the combined (red dashed lines) and individual 
(primary in dashed-dotted green, secondary in dashed-dashed-dotted
blue) fits of the two binary components.}
\label{atlas_binary}
\end{figure}

To be able to extract the longitudinal field value from the LSD profiles with
the center-of-gravity method, we first needed to separate the components of the
multiple system. We fitted each LSD Stokes I profile with 2 gaussian components
(see Fig.~\ref{atlas_binary}) to account for the two main components. The third
weaker component of the system might be visible on the wings of the intensity
profiles, but is neglected here. We then substracted the fit of the broad
component from the observed I profile, and we use the resulting LSD I profile of
the narrow component only to derive the magnetic field value. This method is
described in more details in \cite{neiner_pacwb}. The $B_l$ values are computed
with a mean wavelength of 500 nm and a mean Land\'e factor of $\sim$1.5, over an
integration range centered on the line of the magnetic component (as defined by
the binary fit), i.e. 32.6 and 46.2 km~s$^{-1}$ for the two measurements
respectively, and spanning $\pm$62 km~s$^{-1}$ (see Fig.~\ref{atlas}). The
significance level of the $B_l$ measurements is high, while the one of the N
measurements is very low, which confirms that the signature is of stellar
magnetic origin. Longitudinal field results are shown in
Table~\ref{tableresults}. 

Note that four archival spectropolarimetric measurements of Atlas exist,
obtained with the Musicos spectropolarimeter, which equipped TBL before Narval
was installed. In his Master thesis, \cite{silvester} showed than one of these
four measurements exhibits a Zeeman signature as well, in spite of the 240-420 G
error bars. His tentative result is thus confirmed by the much more precise
Narval observations presented here.

\section{Conclusions}\label{discus}

In this paper, we present the discovery of two new magnetic B stars. The
measured longitudinal field values indicate that their polar field strength must
be of the order of 1 kG for i\,Car and 2 kG for Atlas. Moreover, their Zeeman
signatures change from one observation to the next, which indicates that the
magnetic axis is not aligned with the rotation axis. While the B8V component of
Atlas shows very simple Stokes V signatures typical of a dipolar field, as
observed in most magnetic hot stars \citep{grunhutneiner2015}, the very high S/N
of the Stokes V profiles of the B3V star i\,Car allows us to see possible
additional weak bumps in the wings of the line (around -30 and 50 km~s$^{-1}$),
suggestive of a more complex field. Only a handfull of non-dipolar magnetic hot
stars are known as of today: HD\,37776, HD\,32633, HD\,133880, HD\,137508, and
$\tau$\,Sco.

The two new magnetic main-sequence B stars presented in this Letter are very
bright (V$<$4), and are thus ideal for multi-technique studies. For example,
they are bright enough to be observed with flux-demanding techniques, such as
interferometry. In particular, they can be observed by the BRITE constellation
of nano-satellites, making it possible to detect spots and probable pulsations,
and to perform seismology. Combining magnetic and seismic information is the
only way to probe the impact of magnetism on the physics of non-standard mixing
processes inside hot stars. For example, if the strength of the magnetic field
is sufficient \citep[see e.g.][]{zahn2011}, it inhibits mixing inside the star
and thus allows us to constraint the amount of overshooting needed in seismic
models. Second, the magnetic field produces a splitting of the pulsation modes
and a change of their amplitude. Knowing that a field is present, its strength,
and its  configuration  allows us to securely identify the pulsation modes and
to put strong constraints on seismic models. In addition, combining magnetic and
seismic information allows us to probe the internal magnetic field, while
spectropolarimetry alone only provides information about the surface field.

Furthermore, Atlas is a member of the Pleiades and, therefore, probably the
brightest upper main sequence magnetic star whose age is known \citep[$\sim$100
Myr,][]{soderblom2009}. Moreover, the distance and proper motions of i\,Car make
it a possible member of the younger Lower Centaurus-Crux group of the Sco OB2
association \citep[$\sim$10 Myr,][]{degeus1989,dezeeuw1999}. Knowing stellar
ages of magnetic stars is interesting for fossil field and stellar evolution
studies, and is a very strong asset for asteroseismic modelling. 

However, additional spectropolarimetric observations are required to
characterise the discovered magnetic fields in detail, before precise
constraints can be provided for seismic models. In particular, it is necessary
to determine their polar field strength, the obliquity of their magnetic axis
with respect to their rotation axis, and the possible multipolar components if
the field is not a pure dipole. Follow-up observations of these two targets are
already scheduled with Narval and HarpsPol, and their detailed characterization
and modelling will be the purpose of a future work.

Atlas and i\,Car are among the very few bright (V<4) magnetic B stars discovered
as of today. Their study will undoubtly provide critical information about the
physics inside hot stars. This will, in turn, be of great interest for other
aspects of stellar physics, in particular stellar evolution, since magnetic
fields influence mass-loss and angular momentum as the stars evolves.

\section*{Acknowledgements}

CN thanks James Silvester and Gregg Wade for communicating their Musicos results
for Atlas, and the referee, John Landstreet, for his insightful comments. CN and
AB acknowledge support from the ANR (Agence Nationale de la Recherche) project
Imagine. This research has made use of the SIMBAD database operated at CDS,
Strasbourg (France), and of NASA's Astrophysics Data System (ADS). 


\bibliographystyle{mnras}
\bibliography{articles} 


\bsp	
\label{lastpage}
\end{document}